# Adaptive Traffic Signal Control for Developing Countries Using Fused Parameters Derived from Crowd-Source Data

**Sumit Mishra[a], Vishal Singh[b], Ankit Gupta[c], Devanjan Bhattacharya[d,e], Abhisek Mudgal[c]**

[a]The Robotics Program, Korea Advanced Institute of Science and Technology, Daejeon, Republic of Korea
[b]Member of Technical Staff, VMware Inc, Palo Alto, California, USA
[c]Department of Civil Engineering, Indian Institute of Technology (Banaras Hindu University), Varanasi, India
[d]Marie Skłodowska-Curie Actions TRAIN@ED Fellow, School of Law & School of Informatics, University of Edinburgh, United Kingdom
[e]NOVA Information Management School (NOVA IMS), Universidade Nova de Lisboa, Campus de Campolide, 1070-312 Lisboa, Portugal
sumitmishra209@gmail.com, sivishal@vmware.com, ankit.civ@iitbhu.ac.in, d.bhattacharya@ed.ac.uk, abhisek.civ@iitbhu.ac.in

**Abstract**

Advancement of mobile technologies has enabled economical collection, storage, processing, and sharing of traffic data. These data are made accessible to intended users through various application program interfaces (API) and can be used to recognize and mitigate congestion in real time. In this paper, quantitative (time of arrival) and qualitative (color-coded congestion levels) data were acquired from the Google traffic APIs. New parameters that reflect heterogeneous traffic conditions were defined and utilized for real-time control of traffic signals while maintaining the green-to-red time ratio. The proposed method utilizes a congestion-avoiding principle commonly used in computer networking. Adaptive congestion levels were observed on three different intersections of Delhi (India), in peak hours. It showed good variation, hence sensitive for the control algorithm to act efficiently. Also, simulation study establishes that proposed control algorithm decreases waiting time and congestion. The proposed method provides an inexpensive alternative for traffic sensing and tracking technologies.

## 1. Introduction

Traffic congestion leads to loss of fuel and productive time, and increase in driver stress, air and noise pollution. In developing countries, excessive traffic demand and unplanned growth of city road networks worsens the situation. Lack of appropriate mass transportation and lane free traffic with lack of integration makes the traffic more chaotic [1]. In a chaotic traffic, driver stress from direct and indirect losses may lead to an increase in risky driving such as red light running or speeding that may end up in a crash. Red light running (RLR), which occurs when a vehicle enters an intersection during the red time, accounts for 16% to 20% of crashes at the urban signalized intersection in four states in the U.S.A [2]. In addition, in developing countries, vehicles entering an intersection do not maintain lane discipline. Driver anxiety affected by the wrong management of traffic lights may lead to these crashes [3].

Effective management of the road network is required to mitigate congestion and related problems. Road network capacity enhancement may not be an appropriate solution. However, suitable route/traffic management that ensures uniform distribution of traffic in a road network in space and time is likely to be a more realistic solution to congestion [4]. For reducing traffic congestion, a framework that integrates dynamic vehicles rerouting and traffic light control has been proposed [5]. However, drivers are less inclined to change the usual routes to their destination. If they are forced to take alternative routes, chances of crashes may increase if drivers are less familiar with alternative routes [6, 9]. Hence along with facilitating smart mobility, optimal signal time management is needed.

Most traffic lights in developing countries have fixed cycle length and signal timings that are not revised for years. Due to fluctuation in traffic demand, it is not optimal to have fixed signal timing [14] throughout the day. A possible solution can be to use historic data and allocate different timings for different hours (say) of the day [8]. While it takes only five minutes to change traffic from free flow to congestion, dissipation takes a long time [10, 11]. The higher the degree of congestion, the longer it would take to dissipate it. At free flow conditions, traffic flow varies linearly with traffic density. However, after the peak flow is reached, further increase in density degrades the traffic state that is highly uncertain. It may be difficult to revert to the previous state as described by the fundamental diagram of traffic flow. Therefore, different signal timings at different times of the day may not be an optimal solution especially for non-recurrent congestion. If a short burst of traffic enters a road link that is operating at capacity, and temporarily pushes the traffic density up, the flow rate will drop below maximum. This decreased flow rate will further increase traffic density. This domino effect leads to rapid flow rate decay that will culminate in traffic collapse and jam. An effective solution to this problem may be to control the upstream demand dynamically based on the severity of congestion [9]. Having the traffic to operate at an optimal level may not be the right approach as a slight increase in demand or speed drop may result in severe jam. Signal timing that results in suboptimal flow may be more desirable as there would be some room for increase in demand or speed drop.

Traditionally, sensors are needed for extracting traffic state parameters. Rather than investing in sensor deployment and maintenance, it may be more economical to use crowdsourced data in consortium with limited sensor data [13, 35, 56]. Crowdsourced data accessed through an application program interface (API) provides close to real-time traffic conditions. Traffic management officials of major cities in India, like Delhi and Bangalore, intended to manage road congestion by adjusting cycle time on hourly basis rather than considering real-time traffic status. However, in general, this solution did not lead to a decrease in traffic breakdown or jams. In the proposed method, congestion is tracked at a resolution of cycle length rather than on an hourly basis based on data available through APIs.

Our goal is to design an optimal signaling method for lane free traffic that uses real-time traffic data while preventing traffic from entering the critical region beyond capacity. Fully adaptive traffic control systems may not be an effective solution for developing countries because of lack of funds as well as challenges associated with implementation on unplanned road networks [12]. Heterogeneous traffic with limited lane discipline makes it even harder to apply classical methods of signal time computation. For inappropriately planned roads, initial cycle time and split time ratio must be properly estimated. This entails the need of optimizing signal time at an intersection based on the congestion level, without disturbing the green to red time ratio. To this end, the proposed method adapts cycle length based on existing congestion level while maintaining the red to green time ratio. This constraint may make computation simpler which is desirable especially when low-cost controllers are to be used. For adjusting the traffic signal timing, a simple algorithm is proposed that leverages the principle of 'additive increase multiplicative decrease' (AIMD) of Transmission Control Protocol/Internet Protocol (TCP/IP) congestion control protocol of packet data. The AIMD algorithm is a feedback control algorithm that initially operates by allowing linear (*additive increase*) growth of available resource (green time) up to a level slightly less than maximum resource when the *multiplicative decrease* component is triggered and thus resource is slashed. In an AIMD control strategy, flows from multiple approaches of an intersection will eventually converge to use equal amounts of a shared available resource (green time in case of signal). Based on an analytical and simulation-based study on a simple 4-way intersection, the proposed method enables the system to sustain different traffic intensities peaks while enhancing the capacity of the road network during congestion.

In the proposed work, adaptive control is applied to each traffic intersection independently. Generally, for optimizing traffic flow of a well-organized, lane-bounded road traffic, signal coordination is crucial. However, in a heterogeneous traffic stream, there is a significant variation in driver behavior for vehicles ranging from two wheelers to large trucks. Vehicle heterogeneity and lack of lane-discipline makes it difficult to predict events at an intersection (microscopic level). It also creates erratic traffic during short intervals leading to phantom jams [15]. This makes it quite challenging to implement signal coordination in developing countries. Signal coordination [5] has been tried before by letting some additional traffic pass by in a green wave, but overall, it had very little positive effect on congestion mitigation at macroscopic or network level. The objective of this work is to propose a technique or framework for controlling the signal cycle time such that it
- Utilizes crowdsource data rather than expensive sensor deployment
- Requires minimum processing that can be performed locally using low-cost microcontrollers
- Allows the operators to override parameters (such as keeping the green to cycle time ratio constant) thereby making the system more flexible

Rest of the paper is outlined as follows. Section 2 outlines past studies on road monitoring techniques using sensors and various algorithms based on crowdsourced data, API, and social networks. This section concludes that a system is needed that is not infrastructure intensive yet efficient. Section 3 provides reasoning behind using the AIMD principle and describes the proposed methodology and system architecture. Section 4 analyses and explains the programming involved in the proposed signal timing strategy. In Section 5, validation of the proposed framework is presented through a simulation study. Section 6 presents the conclusions. The overall flow of the main components of this manuscript are depicted in Fig. 1.

## 2. State of the Art and Real-Time Traffic Congestion Management for Developing Countries

Real-time traffic information such as mean speed, traffic density, travel time, delay, congestion level, queue length, are vital for traffic management. Some parameters are directly measured while others are computed from the raw data. Traditionally, such traffic information has been collected using presence of speed sensors as shown in Table 1. The vehicle sensing infrastructure comprises inductive loop detectors, magnetic sensors, LIDAR, RADAR and video cameras [16-18]. Sensors like induction loop, magnetic sensor can provide only point specific data however sensors like camera, LIDAR, RADAR can be used for vehicle trajectory extraction and analysis. Modern approaches for collecting traffic information involve wireless sensor networks (WSNs) [19, 20] RFIDs [21], or vehicular communications (VCs) [22]. Some advanced Intelligent Transportation System (ITS) and schemes obtain highly precise information from a mesh of sensors and dedicated infrastructures. The WSN-based systems rely on numerous roadside sensors to calculate direct and indirect traffic related parameters.

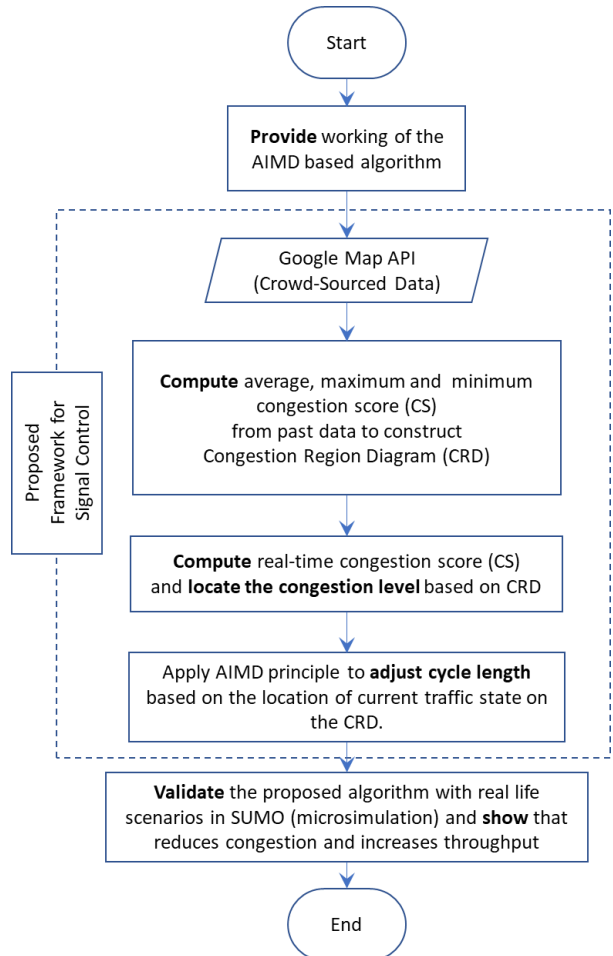

Fig. 1. Overall flow components comprising the methodology

A full-fledged ITS is a perfectly connected network of different technologies, like WSN, CCTV cameras, vehicle onboard sensors, road sensors that lead to more reliable identification of congestion precursors. Several levels of systems ranging from Level-0 to Level-4 have evolved over time. Level-0 includes TRANSYT (1969, U.K.) which is a fixed time and actuated control. At Level-1 there is SCATS (1979, Australia) that incorporates central control and off-line optimization. Then, at Level-2 SCOOT (1981, U.K.) was introduced. This has a central control and on-line optimization. RHODES (1992, U.S.A.), a Level-3 system, is based on distributed control. Finally, MARLIN-ATSC (2011, CANADA) is a Level-4 technology that integrates distributed self-learning [48]. Other systems for signal management and control are ACS-Lite (Adaptive Control Software Lite), SPOT / UTOPIA (Urban Traffic Optimization by Integrated Automation), MOTION (Method for the Optimization of Traffic Signals in On-line controlled Networks), ITACA (Intelligent Adaptive Control Area) and RTACL (Real-time Traffic Adaptive Control Logic). An indigenous developed system CoSiCoSt (Composite Signal Control Strategy) is targeted to cater the typical Indian driving and traffic conditions such as non-lane based driving in mixed traffic flow conditions. CoSiCoSt still needs vehicle sensing infrastructure like induction loop and virtual loop as used in computer vision [31, 32]. Recently, advanced traffic signal control algorithms, based on evolutionary [59, 60] and reinforcement learning [61], are being leveraged. However, these algorithms are computation expensive so using traditional control [62], signaling control techniques can be devised. Similarly, in the present work AIMD based cheap computation control technique is adopted.

On the infrastructure side, ITS uses information and communication technology (ICT) to streamline the operation of vehicles, manage vehicle traffic, assist drivers with safety and other information [26, 27]. Connected vehicles refer to connections with different internal and external environments, i.e. supporting interactions of vehicle-to-sensor on-board (V2S), vehicle-to-vehicle (V2V), vehicle-to-infrastructure; road (V2R) and vehicle communicating directly to the internet (V2I) [28]. There are other Infrastructure-Free Vehicular Networks (V2V) that use these technologies [26, 28, 29] and more accurate traffic related data (individual vehicle trajectory) can be obtained through these [30]. Deploying a full-fledged ITS infrastructure is expensive even for high income nations. Also, Delhi (India) does not have these sensing infrastructures. At limited places there are traffic sensors, however, they are not sufficient for adaptive signaling control as they lack minimum spatial coverage and access to data is not easy. In Delhi at most intersections, cycle time for different traffic scenarios is estimated via historic data and CoSiCoSt and deployed according to time of the day plan. Further, initially deployed systems were not effective due to lack of real-time data. Different techniques [24] can be used for acquiring traffic data for traffic management purposes are shown in Table 2.

Table 1 Different sensors used for traffic sensing

| Sensors | Usability |
|---|---|
| Inductive loops | Count, Presence, Speed estimation (single lane) |
| Cameras | Count, Presence, Speed estimation and classification (multiple lane) |
| Magnetometers | Count, Presence, Speed estimation (using set of 2 sensors) and classification (single lane) |
| Acoustic sensors | Presence, speed estimation and classification (multiple lane) |
| Radar/LIDARs | Count, Presence, Speed estimation and classification |
| Accelerometers | Vehicle speed, wheelbase and distance |
| Infrared Sensors | Count, Presence, Speed estimation and classification |
| Ultrasonic Sensors | Count and presence (multiple lanes) |
| Bluetooth and Wi-Fi Sensor | Count, Presence, Speed estimation (using set of 2 sensors) |
| Wireless Communication Devices (Sensing via received signal strength) | Presence, Count of vehicle for single Tx-Rx pair, Speed estimation using multiple Tx-Rx pair |

Table 2 Different technique for road traffic data grabbing in addition to data from sensors directly

| Crowdsourced data | Statistical Analysis | Sensors and Crowdsourced data |
|---|---|---|
| 1. Direct fleet data access via road public transport or private cab services, if agreed. Data source comprise of sensors that can be on-board vehicle, like Cameras, Acoustic sensors, Radar/LIDAR, Accelerometer. Vehicle specific characteristics can be calculated for vehicles which have sensors onboard. Further, stretch specific characteristics like average speed and queue length can also be calculated. | Data is historical or simulated | Fusion of direct crowdsourced data and sensor data through appropriate techniques is possible for accurate statistical analysis. |
| 2. Access via 3rd-party like Google Maps. Data source in the form of new empirical parameters is provided by 3rd-party eg. CI (Congestion Index), ETA etc. Stretch specific characteristics like average speed, congestion level can only be calculated for traffic intensity categorization. | Stretch specific characteristics can be calculated based on different sensor, survey and crowdsourced data availability. | After appropriate sensor and 3rd party crowdsourced data fusion, vehicle specific characteristics can be calculated for vehicles which have sensors onboard. Further, stretch specific characteristics like average speed can also be calculated. |

Some studies promote the use of centralized systems with or without hardware [36-39]. However, setting up a new system especially that is not calibrated for heterogeneous traffic and driver behaviour would likely not provide optimum results [40]. It may be better to use a system that leverages data from platforms already prevalent in the market and have a large user base.

For example, companies such as Google and Bing [13] collect data using crowdsourcing and make them available via APIs while ensuring user's security and privacy. Information is presented via APIs in different forms as per visualization needs. Implementing adaptive signal control using crowdsourced data may need fusion of data from various sources [35]. The way congestion was quantified traditionally may be replaced by new parameters based on the availability as well as richness of crowdsourced data.

Processing raw crowdsourced data requires huge processor resources, data storage systems, dedicated personnel and a good user base. Big companies like Google and Bing have developed efficient and reliable ways of processing and managing location data transmitted by a large number of cellphone users that use their services and devices. Whenever available, data from local highway authorities, public transport and taxi fleets are also integrated with crowdsourcing data to increase the accuracy and reliability of estimated parameters (such as average speed and travel time). Some data have been collected using survey vehicles like the Google Street View cars which have gathered 7 million miles of road infrastructure data in the USA. There are other sources of data as discussed by many Google Maps employees [49]. However, their efficiencies vary from place to place [50] and data accuracy depends on time, place, algorithms among other factors. The limitations will fade away as adoption of Google's mapping services and smartphones increases. Nair et al. (2019) compared Google API and loop detector data and found RMSE deviation within 11 km/h for 90 percent of 53 test locations. In addition, speed obtained from Google API was within 2 km/h of the speed estimated using the floating car method [34]. Raw crowdsourced data is processed to estimate live traffic status as color coded road sections along with estimated time to arrival (ETA) from origin to destination. Color coding is qualitative while ETA is quantitative presentation of traffic speed. For the color representation model, Bing provides TrafficLayer class (deprecated) and TrafficManager [46] class for getting live traffic status. The TrafficLayer class has many inbuilt functions like *hide*, *show*, *showFlow* etc. With the help of these functions, we can quantify the traffic congestion level at desired locations at a particular zoom of a Google Map tile in the form of color codes. Google maps also provide JavaScript APIs, which have a variety of functionalities useful for map editing. One of these is Traffic Layer API [47], which provides options for live traffic in four colors (Fig. 2). ETA is also provided by both Google and Microsoft [46] through 'distance matrix API'. These APIs differ in terms of coverage and functionality. Google has good coverage in developing countries due to high number of Android smartphone users. Also, Google ensures and approves accessing congestion data on the map for instructional or illustrative purposes [23].

As an alternative to investing in sensor infrastructure, some studies have proposed to use distributed computation power and crowdsourced traffic data [35, 41, 42]. Others suggest leveraging data from social networks especially twitter [43, 44]. These studies do not incorporate derivation of parameters needed for real time congestion tracking. It may be feasible to use infrastructure-free techniques that take advantage of large amounts of crowdsourced data to monitor congestion levels. The APIs data can't be used directly as none of the API publishes live average congestion status. The crawling speeds in the urban environment are indicative of congestion [45] as evident by API data like ETA data and color tracks of speed ranges. There is a need to map APIs data to a parameter representative of average congestion in a desired area. In [35] an empirical parameter 'CI' (Congestion Index) is adopted considering ETA data only. However, in addition to ETA data, the color code information as it represents the traffic state, incorporates other information that can be important to quantify the congestion level more realistically. Few researchers [56] used only the distance matrix API model for computing four cycle times based on congestion level of intersection without considering the proportion of congestion contributed by different approaches in the past (historical congestion data). This technique is implemented in Hyderabad, India [57, 25]. Researchers [58] used segment colors from Google API for comparative study of vehicular pollution and policies. We propose a set of empirical parameters for congestion status modeling of an intersection by fusing the techniques mentioned in the above models [57, 58]. We propose to quantify congestion based on real-time ETA (quantitative) and color code information (qualitative) as available for each approach at an intersection.

According to robust data needs, a user may use any combination of model and API service from different companies to compute congestion levels based on crowd-sourced data. This will not only complement information from a variety of sources but also provide optimal information. Optimal information is gathered by avoiding Nash equilibrium point which occurs because independent services (traffic congestion informative apps) compete for system resources independently while being unaware of the characteristics of other services [51]. Therefore, if congestion parameters are estimated using data from several sources and services (Google and Bing etc.), it would more likely give rise to optimal conditions [25, 34, 35, 41-44, 56].

## 3. Proposed Methodology and Architecture

The proposed algorithm computes the congestion score (CS) for an intersection every cycle length and compares it with the historically average CS to produce a qualitative measure of congestion level (no congestion, low congestion, mild congestion and severe congestion) for the given intersection. Then, based on the existing congestion level, cycle length is updated using the principle of AIMD which is commonly used for managing data packets in TCP/IP protocol of the internet [52].

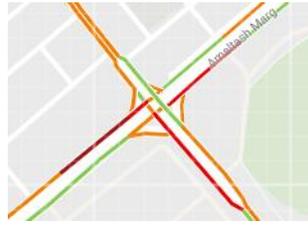

Fig. 2. Color coded tracks from Google Traffic API based on different speed ranges

### 3.1 Signal Optimization Using the AIMD Principle

AIMD turns out to be a powerful, effective and simple mechanism for congestion control. The problem of congestion control involves conflicting requirements such as maximizing the intersection throughput on all approaches. Every approach must get a fair share of green signal time. Therefore, the idea is to use an algorithm for congestion control that has the approach of operating close to maximum utilization by maintaining the fair share. For maximum utilization, the system will converge to a point where every intersection link (assuming everything else is equal) will get approximately '1/m' traffic density of the whole 4-way infrastructure density, if there are 'm' links at the intersection. Since it provides the best optimum allocation of time by spreading traffic equally, traffic jams are minimized.

The concept of AIMD can be understood graphically using the so-called "CHIU JAIN" plot [52]. Explanation is provided for two approaches (with flow rate A and B respectively) or links that are competing for resources at the intersections. The following analysis can be done with any number of intersection links using a higher dimension vector space. Traffic flow is directly proportional to green time. Flow rate A and B are plotted on the X-axis and Y-axis respectively. Given the overall intersection capacity as 'C', the diagonal line "A + B = C" represents efficient operation while the dotted line "A = B" depicts fairness. The region above the diagonal line where A + B < C represents underutilization of the traffic signal while the region below it depicts overutilization of the traffic signal. For a fair network, A will be equal to B. Hence, the flow values should fall on line "A = B" after stabilization (using AIMD algorithm). Now, to maximize overall link utilization the two links must operate at total capacity equal to A + B or on the line (A + B = C). The AIMD algorithm aims to achieve flow A and flow B by first increasing additively and then decreasing multiplicatively (actually by dividing) as it crosses the line (A + B = C). Figure 3 also depicts the behavior of AIMD through a series of cycle lengths namely T1 through T4. For cycle length T1 flow B is operating well above its fair share and flow A is operating well below its fair share but the overall flow is less than 'C'. So, both flows are in an additive-increase mode with equal increments. If both of these are increased additively, the point must move parallel to line A = B until it crosses line A + B = C. At that point, the network becomes overloaded. Hence, a multiplicative decrease of a factor, say '1/2', is applied. After that, the system goes back to an additive increase mode again. Now, since the 'multiplicative decrease' decreases B more than A (it is a multiplicative factor), the traffic state approaches the fair line (A=B). The point T3 is closer to the fair line than T1 and T3. Over time, the system oscillates between overloaded and under loaded conditions pushing the network until it's just slightly overloaded, and then back off a little bit. This scaling with the "multiplicative decrease" factor causes the flow to converge towards the desired equilibrium point i.e., the green dot (intersection of A = B and A + B = C). At that point, the system is fair to both the approaches while also utilizing the network efficiently. Over time, AIMD causes a set of flows to achieve both of the desired properties (fairness and efficiency).

To deploy an AIMD principle-based algorithm for adjusting the signal timings, microcontrollers (as slaves) are needed for each lane. These slaves are connected to the central microcontroller unit server (MCU) which acts as the master. Generally, various ready-to-use traffic-light hardware boards (master and slave) are available and are deployed at various traffic-lights. The proposed method would use these microcontrollers to deploy the AIMD algorithm to control signal timings based on the level of congestion. Further, the central MCU is wirelessly connected to the traffic control center where processing may be done. The MCU at the intersection receives the timing according to the congestion level and then communicates it to every traffic light via a small network as shown in Fig. 4. If a standalone solution is needed for each intersection, then the central MCU has to be powerful. Otherwise, a Single Board Computer (SBC) like Intel NUC can independently perform the functionality of a server desktop PC housed at the traffic control center. Further, traffic light control hardware implementations are susceptible to threats like security, system crash due to bad internet connection or overheating. For dealing with these threats a reliable operating system crash susceptible architecture must be used [30].

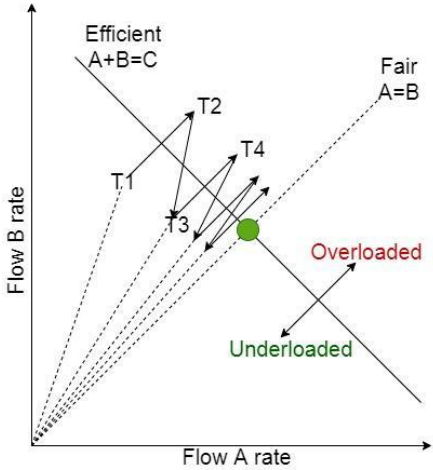

Fig. 3. Chiu-Jain plot

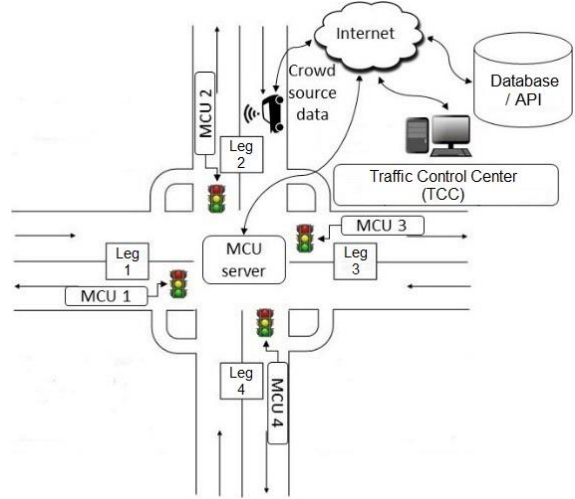

Fig. 4. Interconnected microcontroller network

### 3.2 Development of Adaptive Signal Timing and Control Algorithm

Congestion score (CS) of a given intersection is computed using real-time as well as historically available qualitative (color) and quantitative data (ETA) for traffic on road sections spanning from the geometric center of the road intersection to 500 m (or a suitably calibrated distance along the intersection legs) on all approaches to the intersection. The ETA data is obtained from distance matrix API while color codes data is extracted from Java map API. Congestion score is the product of color-based congestion measure ($CM_{color}$) and ETA based congestion measure $CM_{ETA}$ (See Eq.1 to Eq. 4). The average congestion score $CS_{avg}$ is computed as the product of average $CM_{color}$ and average $CM_{ETA}$ (See Eq. (5)). Average $CM_{color}$ or $CM_{color,avg}$ is computed as the hourly mean of $CM_{color}$ recorded during the same hour last week while average $CM_{ETA}$ or $CM_{ETA,avg}$ is computed as long term (say monthly) mean of $CM_{ETA}$. The color-based congestion measure is computed for the whole intersection while ETA based congestion measure is real time ETA weighted by past ETA values observed on the given link. The proportion of congestion or weight is given by $w_i = LETA_i/\sum_{i=1}^{m} LETA_i$, where $m$ is the number of links. The congestion regions (Region-1: no congestion, Region-2: low congestion, Region-3: moderate congestion, Region-4: severe congestion) are defined based on three congestion scores namely $(CS_{min} + CS_{avg})/2$, $CS_{avg}$ and $(CS_{max} + CS_{avg})/2$ as shown in Fig. 5. Here, $CS_{min}$ and $CS_{max}$ are the minimum and maximum $CS$ observed during the same hour last week. Congestion regions are defined for each hour. $CS$ is calculated using crowdsource data only. If the crowdsource data is not reliable and more stable data from sensors is available then sensor data needs to be converted to ETA to compute all the parameters.

As mentioned earlier the Google API does not publish congestion levels in the digital format explicitly. The distance matrix API provides 'duration_in_traffic' which is real-time ETA for each link (origin-destination pair). The distance matrix API takes a pair of original and destination (in terms of latitude and longitude) points anywhere on the road and returns the estimated time of arrival (ETA) between those two points. For a four-legged intersection with both way streets on all four approaches, there are eight links. ETA is basically the estimated travel time between the identified origin and destination and is given by the distance between the origin and destination divided by the real-time speed between the pair of points. The API denotes ETA as 'duration_in_traffic'. The API also provides 'duration' which is averaged ETA over 'n' samples for long time periods (say months) and is referred to as 'LETA' as shown in (Eq. 5). LETA values are obtained for each link (say m in this case). The color information is converted to a numeric value called $CM_{color}$ which is computed as a weighted average of green, orange, red and dark brown color fractions, $f_c$. The weights are chosen as 0.25, 0.5, 0.75 and 1.00 for green, orange, red and dark brown color fractions respectively as shown in Eq. 2. Color fraction for a color is the proportion of pixels of that color. The given weights were chosen to map congestion, a non-linear phenomenon, to a linear scale. $CM_{color}$ varies from 0.25 (for all green pixels) to 1.00 (for all red pixels) and is higher for higher congestion levels.

Capturing and processing images may be efficiently done in the background through various virtual machine software tools. Also, a single image can provide traffic state data of all the approaches in a single Java map API. There is no need to make a distance matrix API call that is more data intensive (need for linkwise origin/destination data). These would make the overall process more efficient. In summary, the following three steps are required:

1) An image of 'Google Map API' tile centered at the center of the intersection is captured at a suitable zoom.
2) Clutters like landmarks, parks, local roads, water bodies, buildings and other irrelevant items are removed to increase the speed and accuracy of image processing.
3) Image processing is done to calculate the fraction of pixels of different colors.

A four-legged intersection with all bi-directional approaches comprises eight intersection links that have combined influence on the resultant congestion experienced at the intersections. $CM_{ETA}$ is computed as a weighted sum of ETA on each link. The weight, $w_i$ is the proportion of congestion (depicted by LETA) contributed by a link historically.

$$CS = CM_{color} * CM_{ETA} \tag{1}$$
$$CM_{color} = 0.25*f_{green} + 0.50*f_{orange} + 0.75*f_{red} + 1.00*f_{dark\ brown} \tag{2}$$
$$CM_{ETA} = \sum_{i=1}^{m}(w_i * ETA_i) \tag{3}$$
$$CS = (0.25 * f_{green} + 0.50 * f_{orange} + 0.75 * f_{red} + f_{dark\ brown}) * (\sum_{i=1}^{m} w_i * ETA_i) \tag{4}$$
$$CS_{avg} = CM_{color,avg} * CM_{ETA,avg} = CM_{color,avg} * \sum_{i=1}^{m}(w_i * LETA_i) \tag{5}$$

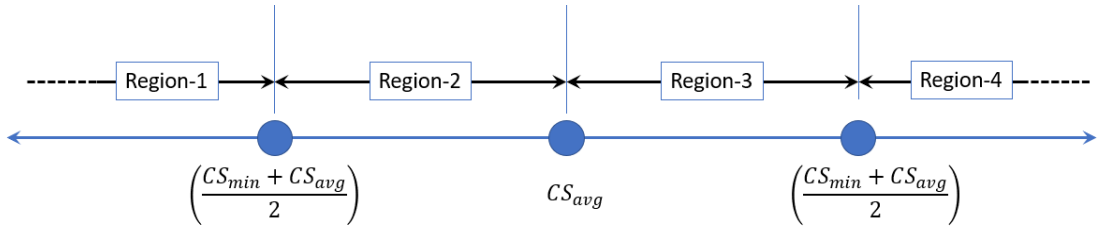

Fig. 5. Congestion Regions Diagram (CRD) based on congestion score

Quantification of congestion level is an empirical scheme which the authorities may wish to define using well-established parameters like cruising speed, queue length and degree of saturation at an intersection. These parameters may be estimated using traffic flow models using the available crowd-sourced data (e.g. ETA). For example, average speed can be estimated using travel time (ETA) and the length of the link for which congestion level is to be defined. In order to calibrate the parameters using a simulator (e.g. SUMO), the estimated average speed (on all links) may be compared with the field values. Once calibration is done and simulation model of the intersection is established, other parameters may also be obtained. Different schemes like a heuristic algorithm as described in [4] and an evolutionary algorithm as described in [7] may be adopted for creating real-time simulation trajectories that can be used to quantify the congestion level. When available, the sensor data can be integrated into the simulation step which can reduce the error further [30]. The fusion of simulated and recorded data may make quantification of congestion more accurate.

In the proposed framework, cycle length $T$ is adjusted based on the level of congestion experienced by the intersection. The aforementioned signal timing control algorithm is depicted in Fig. 6. Maximum cycle length $T_{max}$ for a given intersection can be found by using CoSiCoSt or empirical data. For an intersection with $m$ approaches the cycle length $T = \sum_{i=1}^{m} G_i$ where $G_i$ denotes the durations of green time. The amber/yellow (waiting) time of each split is unaffected by the algorithm and hence ignored in the equation. Since, smaller cycle length is suitable for a low traffic volume network, the algorithm proposes to increase the cycle length when congestion (higher volume) hits the intersection. Also, from Webster formulae it can be seen that cycles length in the range of (0.75 - 1.5) times the optimum-cycle-length do not significantly increase delay [53]. The proposed algorithm keeps track of the congestion level and adjusts the cycle length within the optimal range.

The algorithm suggests setting the cycle length to $T_{max}/2$. At Region-4 congestion (severe), cycle length is increased from $T_{max}/2$ to $T_{max}$. The region $T_{max}/2$ to $T_{max}$ lies within (0.75 - 1.5) times the optimum-cycle-length. If the congestion level is maintained at "no congestion" (Region-1), cycle length is kept at $T_{max}/2$. However, as soon as the congestion deteriorates to the next level (Region-2 or worse), the cycle time becomes $T_{max}/2 + $ 'Temp' based on the congestion level. The value of "Temp" is set to $T_{max}/8$, $T_{max}/6$, $T_{max}/4$ and $T_{max}/2$ for Region-1, Region-2, Region-3 and Region-4 congestion levels respectively. It should be noted that 'T/8' is significant as it lies within 15 s to 30 s for cycle time 120 s to 240 s. As congestion worsens, the algorithm keeps adding "Temp" to the cycle length until it reaches $T_{max}$. This is an 'additive increase' in cycle time. After that, it keeps maintaining the cycle time to $T_{max}$ provided the congestion level is the same as the previous congestion level except Region-1. As soon as congestion improves, 'multiplicative decrease' occurs and variable $T$ (cycle length) is set to $T_{max}/2$ and Temp based on the congestion level. This is an iterative process that monitors and adjusts the cycle time. The ratio of green time ($G_i$) and red time ($R_i = T - G_i$) is preserved as decided by the road authorities and therefore the proposed method respects the expertise and computations that went in designing the signal timings of the intersection. This solves the problem caused by a full adaptive control algorithm in traffic patterns and road conditions in developing countries. Also the proposed control algorithm is lightweight and has a time and space complexity of O(1), i.e. constant, both in best and worst case scenario. In each cycle, only fixed number of arithmetic operations are executed.

However, care should be taken to ensure that T/2 is at least equal to the minimum green interval for each movement. Left turns (for right side driving countries like USA, China), minor streets, major streets, usually have different minimum green times with left turns (for right side driving countries) and minor side street intervals are often in the range from 4 s to 10 s while major streets often go higher than 15 seconds [54]. Yellow Clearance, Red Clearance, Walk time, Flashing Don't Walk time should not be changed. Further, the color code data are available in four quantized levels and so to avoid quantization error, four congestion status levels are used.

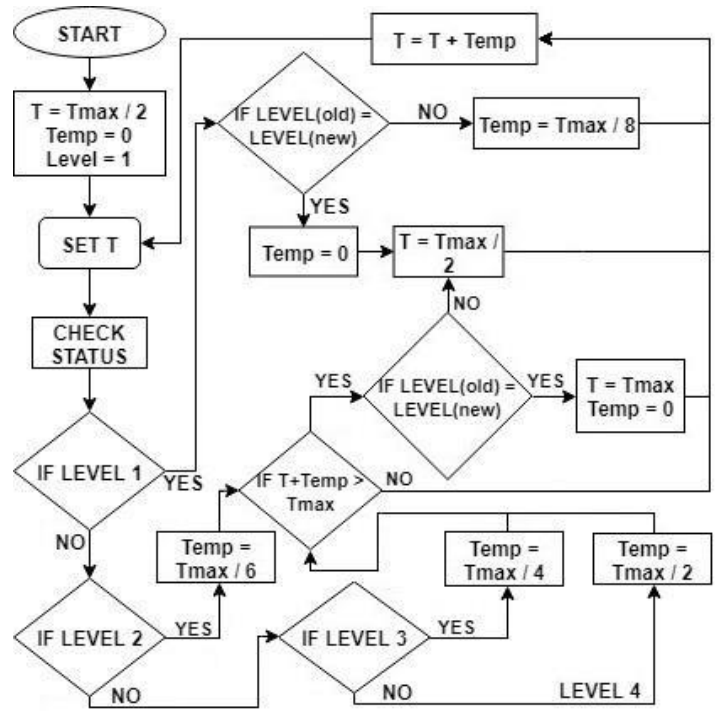

Fig. 6. AIMD based traffic signal timing adjustment algorithm

## 3.3 Programming and Data Extraction

Real-time traffic data was obtained at regular intervals from Google API that allowed access (through API key) to freely available data for research and educational purposes. The distance matrix API request contains many parameters like travel mode, turn restriction, departure time, origin and destination coordinates along with the user's key and information about the required data format (either JSON or XML). The requests return ETA as well as LETA data. For Java map API, an HTML file is generated with the help of Google RoadMapAPI as shown in Fig. 7 (b). Clutters like parks, local roads, water bodies, buildings etc. are turned off using functions in the *StyledMapType* Class. This results in the road network of the study area as shown in Fig. 7 at zoom level 18. Another function in the Google API called *TrafficLayer* is used to obtain real-time traffic in the form of four color codes namely green, orange, red and dark brown which signifies the increasing level congestion. Finally, a processed '.html' file with color codes is obtained. To extract an image from the Google Map, a program (*MapToImage.java*) was developed in Java using swing and I/O packages. The file *MapToImage.java* contains three thread classes (as shown in Fig. 8) which performs their tasks in the following sequence:
- Thread1, Thread2 and Thread3 are initialized simultaneously.
- Thread1 makes an API call and loads the clutter free Google map with color representation of traffic data.

- Thread2 waits for 6 seconds to allow Thread1 to complete loading the processed map, and then takes a screenshot of the computer screen which displays the map. The dimension of the screenshot is 200 mm less than both the height and the width of the monitor. The dimension should be manually set for a given intersection according to preferences. Thread2 saves the screenshot to a '.jpg' format in the same folder as the HTML file.
- Thread3 waits for 7 seconds to allow Thread1 and Thread2 to complete the tasks and then finds the RGB value of each pixel present in the '.jpg' image file. Using a MATLAB code, the images are processed to find the range (Table 3) of RGB values corresponding to the various traffic color codes (green, orange, red and dark brown). Removing the clutters and extracting congestion status from the image introduces an error of about 2% to 3%. This is because (1) the RGB values of a particular pixel do not have a fixed value but fall in a range, and (2) the number of pixels of a particular color for a given picture may vary with the resolution of the graphical processing unit of the image-capturing device. However, the ratio calculated in the proposed method is going to be the same for all devices. The fractions of various colors in the images and subsequently $CS$ are computed by Thread3 using the JSON string produced by the distance matrix API.

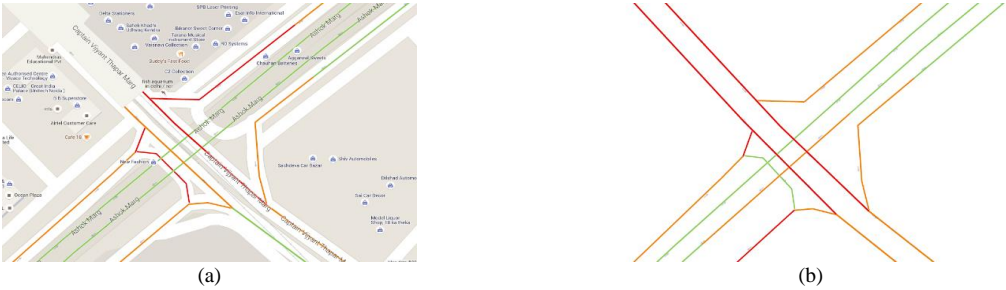

Fig. 7. (a) Images of unprocessed HTML file (b) Images of processed and clutter removed HTML file

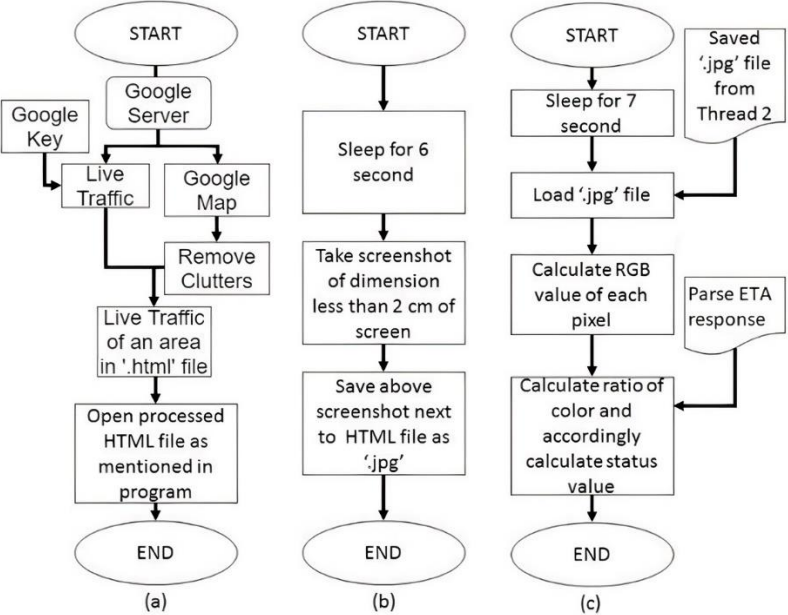

Fig. 8. (a) Thread 1 (b) Thread 2 (c) Thread 3

Table 3 Range for RGB values of different colors of Google maps.

| Color in Google Maps | Ranges for RGB | | |
|---|---|---|---|
| | Red (R) | Green(G) | Blue(B) |
| Green | (125 – 195) | (172 – 232) | (58 – 185) |
| Orange | (180 – 255) | (90 – 190) | (0 – 110) |
| Red | (140 – 220) | (0 – 65) | (0 – 65) |
| Dark brown | (75 – 160) | (0 – 80) | (0 – 90) |

## 4. Proof of Concept of the Proposed Framework

To validate the proposed algorithm, Google traffic data (color codes and ETA matrix) were obtained at three 4-way intersections in Delhi, India from 8 p.m. to 9 p.m. at a frequency of one sample per 2 minutes. Each link consists of three lanes as shown in Fig 9. The proportions of congestion observed on the eight links are shown in Table 4. *CS* is computed and fed into the adaptive signal control algorithm to change the green and red time according to a fixed initial ratio. The frequency distribution of congestion levels [33] observed at these intersections reveals the need for adaptive traffic signal control (Table 5).

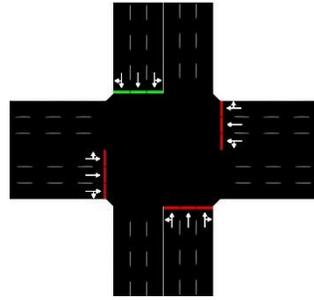

Fig. 9. 4-way Intersection used in the simulation

Table 4 ETA fraction of intersection links (directions and approaches) at different intersections (I).

| Road | Weight(I-1) | Weight(I-2) | Weight(I-3) |
|---|---|---|---|
| 1 | 0.105066 | 0.092925 | 0.09322 |
| 2 | 0.116323 | 0.155227 | 0.084746 |
| 3 | 0.093809 | 0.095037 | 0.112994 |
| 4 | 0.118199 | 0.098205 | 0.112994 |
| 5 | 0.123827 | 0.116156 | 0.135593 |
| 6 | 0.150094 | 0.153115 | 0.144068 |
| 7 | 0.142589 | 0.117212 | 0.152542 |
| 8 | 0.150094 | 0.172122 | 0.163842 |

Table 5 Frequency of congestion levels observed from 8 p.m. to 9 p.m for three intersections (I-1, I-2 and I-3)

| Congestion Levels | Frequency | | |
|---|---|---|---|
| | I-1 | I-2 | I-3 |
| 1 | 6 | 7 | 1 |
| 2 | 2 | 9 | 0 |
| 3 | 14 | 8 | 18 |
| 4 | 8 | 6 | 11 |

For a given intersection, the longer the cycle time the better it is for decongesting the connected links. However, for low congestion, small cycle time is better. The AIMD technique embedded in the proposed algorithm activates whenever congestion changes abruptly. The higher the congestion, the longer it takes to mitigate it. For more severe congestion impulses, a larger additive increase is adopted, while for small impulse, small increment in cycle-time is recommended so that it avoids vehicle queues. The proposed methodology showed that AIMD is applicable to road networks with multiple links. The freely available traffic simulation package SUMO was used to simulate a simple 4-way intersection of two six-lane roads as shown in Fig. 9.

Each lane is 500 m long and the traffic is randomly generated with equal probability in all directions. The traffic includes three types of vehicles i.e. 4-wheeler (cars), 2-wheeler (motorbikes) and buses with probability of occurrence as 0.5, 0.4 and 0.1 respectively. Parameters like maximum speed, acceleration, vehicle length, gap between vehicles and lane change capacity are the inputs to simulation. For detecting congestion, the speed-based detector as discussed in [55] with the cutoff speed of 1 m/s is coded. The detector provides the ratio of number of vehicles with speed less than a cutoff speed on a link to total vehicles on the link. For the purpose of simulation, the four congestion levels are decided by partitioning the range of simulated congestion values obtained by analyzing different simulations for different flow rates.

For practical purposes, the cycle time must not be too large or too small. Road authorities prefer moderate cycle length of 120 to 160 seconds to avoid driver inconvenience and hence red light running events. A cycle time of 120 s to 240 s with 2 s yellow time for each lane is chosen. Fig. 10 shows speed detector data averaged throughout the network for each second of the simulation run for different input (flow) rates of the vehicle in the network. Information about the series depicted in Fig. 10 is presented in Table 6. As observed in the graph of Fig. 10 for low input rate (1 second/vehicle), the small cycle time i.e. Series4, is good as it creates less average congestion and allows a throughput of 4000 vehicles in less time (Table 6). While for high input rate (0.7 second / vehicle), long cycle time i.e. Series3, is suitable for low congestion and throughput of 4000 vehicles in less time (Table 6). This fact is also corroborated by Webster formulae and its more optimized derivatives as studied in [53]. In Fig. 2 of [53], the authors presented graphs of optimum cycle length to intersection flow ratio for different lost time. Those graphs demonstrate that the optimum fix-cycle time increases with the increase in flow.

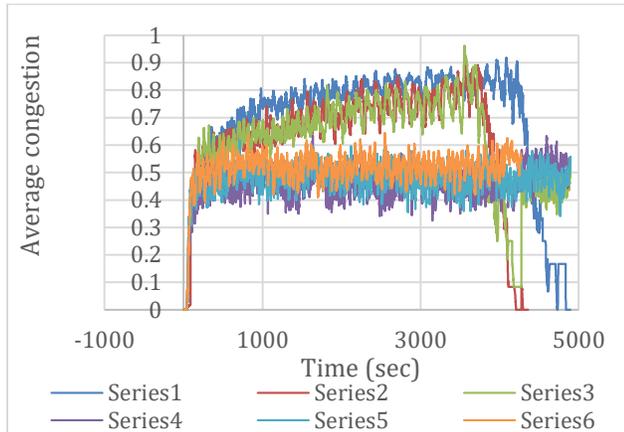

Fig. 10. Comparison of different fix cycle-time with different input rate

Table 6 Different series of Fig. 10

|  | Input vehicle rate (s/veh) | Fix cycle time (s) | Time (s) for 4000 vehicles |
| --- | --- | --- | --- |
| Series1 | 0.7 | 120 | 3613 |
| Series2 | 0.7 | 180 | 3231 |
| Series3 | 0.7 | 240 | 3191 |
| Series4 | 1.0 | 120 | 4160 |
| Series5 | 1.0 | 180 | 4162 |
| Series6 | 1.0 | 240 | 4185 |

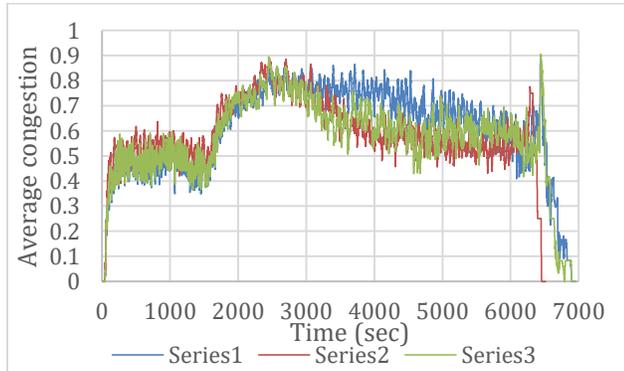

Fig. 11. Comparison of different fix cycle-time with different input rate

Table 7 Different series of Fig. 11

|  | Input vehicle rate (s/veh) for 1500-3000 vehicle no. | Fix cycle time (s) | Time (s) for 6000 vehicles |
| --- | --- | --- | --- |
| Series1 | 0.6 | 120 | 5668 |
| Series2 | 0.6 | 240 | 5449 |
| Series3 | 0.6 | Algorithm | 5650 |

Further we created a simulation in which the input rate is 1 second / vehicle for all, but for vehicles 1500-3000 the flow is 0.6 second / vehicle. This is done to simulate traffic bursts to test the AIMD based algorithm. In Fig. 11, a time series of simulated average congestion is plotted. Three series are plotted in Fig. 11 for which cycle time is decided as shown in Table 7. Initially with the burst, the congestion starts to build up and after sometime when the traffic burst is over, it starts to drop. However, after a congestion burst, the new congestion level is worse than before. From Table 7 and Fig. 11, it can be seen that the throughput of 6000 vehicles is high for Series1 and low for Series2. However, series2 is more congested than Series1 before the traffic burst, which is same as that of conclusion given for low input rate of Fig. 10. Further congestion for Series1 and Series3 is similar before the traffic burst. Series2 and Series3 are equally effective for suppressing congestion as average congestion drops after the peak is almost similar. For Series3, 35 different cycle times were run for 5650 s using throughput volume as 6000 vehicles per hour. Average cycle-time was 163 s, which is much less than 240 s. Hence, the algorithm provides better congestion management not only in normal conditions but also during traffic bursts by maintaining a small average cycle time.

## 5. Conclusions and Future Scope

In this work, we proposed a dynamic traffic control framework that utilizes crowdsourced data (provided by Google Map API) and maximizes intersection throughput in a lane free traffic. In the proposed framework, congestion score is derived by fusing quantitative and qualitative data available from the API. Based on the congestion score, the level of congestion is assessed and this is used to dynamically change the cycle length of the signal using the AIMD principle (TCP/IP internet protocol). The efficacy of the AIMD principle was done using analytical reasoning (Chiu-Jain plot).

Validation of the framework was done by simulating real life scenarios in SUMO, a freely available traffic simulation package. The proposed algorithm decreases congestion by setting small cycle time for low or no congestion situations, and mitigates traffic breakdown similar to that of large cycle time while maintaining small cycle length (on an average) which most drivers would desire.

The method of obtaining the traffic data is infrastructure-free, economically feasible, accurate and scalable for sustainable traffic management in developing countries. The algorithm to compute the congestion status was validated by simulating few intersections in mixed traffic conditions. The proposed algorithm enhances the existing predefined signal timing by reducing congestion and increasing throughput while maintaining the weighted timing ratio for different intersection links. The proposed methodology of controlling traffic signal, can be easily incorporated in smart city framework saving huge amounts of resources like fuel and lost time due to delay. This paper mainly focuses on a holistic framework to optimize intersection throughput utilizing whatever infrastructure free data are available. Further investigation is needed to assess congestion levels in complex road networks that include multiple intersections or grade separated intersections.

## Acknowledgments

The present work in the paper was not funded by any organization. One of the coauthor, Devanjan Bhattacharya has received funding from UKRI ESRC Impact acceleration grant (ES/T50189X/1), and European Union's Horizon 2020 research and innovation programme under the Marie Skłodowska-Curie COFUND grant agreement No. 801215: TRAIN@Ed: 'Transnational Research And Innovation Network At Edinburgh'.

## References

[1] Mahmud, K., Gope, K., & Chowdhury, S. M. R. (2012). Possible causes & solutions of traffic jam and their impact on the economy of Dhaka City. J. Mgmt. & Sustainability, 2, 112.
[2] Mohamedshah, Y. M., Chen, L. W., & Council, F. M. (2000). Association of selected intersection factors with red-light-running crashes. US Department of Transportation, Federal Highway Administration, Research, Development, and Technology, Turner-Fairbank Highway Research Center.
[3] Galatioto, F., Giuffrè, T., Bell, M., Tesoriere, G., & Campisi, T. (2012). Traffic microsimulation model to predict variability of red-light running influenced by traffic light operations in urban area. Procedia-Social and Behavioral Sciences, 53, 871-879.
[4] Zambrano-Martinez, Jorge, et al. "Towards realistic urban traffic experiments using DFROUTER: Heuristic, validation and extensions." Sensors 17.12 (2017): 2921.
[5] Cao, Z., Jiang, S., Zhang, J., & Guo, H. (2017). A unified framework for vehicle rerouting and traffic light control to reduce traffic congestion IEEE Transactions on Intelligent Transportation Systems, 18(7), 1958-1973.
[6] Chakrabarti, U. K., & Parikh, J. K. (2013). Risk-based route evaluation against country-specific criteria of risk tolerability for hazmat transportation through Indian State Highways. Journal of Loss Prevention in the Process Industries, 26(4), 723-736.
[7] Stolfi, Daniel H., and Enrique Alba. "Generating realistic urban traffic flows with evolutionary techniques." Engineering Applications of Artificial Intelligence 75 (2018): 36-47.
[8] Yousef, Khalil M. Ahmad, Ali Shatnawi, and Mohammad Latayfeh. "Intelligent traffic light scheduling technique using calendar-based history information." Future Generation Computer Systems 91 (2019): 124-135.
[9] Mishra, S., Singh, N., & Bhattacharya, D. (2021). Application-Based COVID-19 Micro-Mobility Solution for Safe and Smart Navigation in Pandemics. ISPRS International Journal of Geo-Information, 10(8), 571.

[10] Jain, V., Sharma, A., & Subramanian, L. (2012, March). Road traffic congestion in the developing world. In Proceedings of the 2nd ACM Symposium on Computing for Development (p. 11). ACM.
[11] Geroliminis, N., & Daganzo, C. (2008). Existence of urban-scale macroscopic fundamental diagrams: Some experimental findings. Transportation Research Part B: Methodological, 42(9), 759-770.
[12] Chaitanya Swamy."AUTOMATIC FOR THE PEOPLE" H M, Bangalore Mirror Bureau, Web. 7 Jul. 2017, http://bangaloremirror.indiatimes.com/bangalore/crime/automatic-for-the-people/articleshow/59448220.cms
[13] Eriksson, I. (2019). Towards Integrating Crowdsourced and Official Traffic Data: A study on the integration of data from Waze in traffic management in Stockholm, Sweden.
[14] Mitrovic, N., & Stevanovic, A. (2019). Estimating Peak-Hour Traffic Profiles for Selection of Appropriate Day-of-Year Signal Timing Plans. Transportation Research Record, 0361198119841860.
[15] M. R. Flynn, A. R. Kasimov, J.-C. Nave, R. R. Rosales, and B. Seibold. Self-sustained nonlinear waves in traffic flow. Physical Review E, 2009; 79 (5): 056113 DOI: 10.1103/PhysRevE.79.056113
[16] Sen, R., Cross, A., Vashistha, A., Padmanabhan, V., Cutrell, E., & Thies, W. (2013). Accurate speed and density measurement for road traffic in India. Proceedings Of The 3Rd ACM Symposium On Computing For Development - ACM DEV '13.
[17] Collotta, M., Pau, G., Salerno, V., & Scata, G. (2012). A Novel Road Monitoring Approach Using Wireless Sensor Networks. 2012 Sixth International Conference On Complex, Intelligent, And Software Intensive Systems.
[18] Walton, S., Chen, M., & Ebert, D. (2011). LiveLayer: Real-time Traffic Video Visualisation on Geographical Maps. Retrieved from https://pdfs.semanticscholar.org/929f/aac78137e3b3723994f11f6a82e98fef0ef8.pdf
[19] Pascale, A., Nicoli, M., Deflorio, F., Dalla Chiara, B., & Spagnolini, U. (2012). Wireless sensor networks for traffic management and road safety. IET Intelligent Transport Systems, 6(1), 67.
[20] Collotta, M., Lo Bello, L., & Pau, G. (2015). A novel approach for dynamic traffic lights management based on Wireless Sensor Networks and multiple fuzzy logic controllers. Expert Systems With Applications, 42(13), 5403-5415
[21] Wen, W. (2010) "An intelligent traffic management expert system with RFID technology." Expert Systems with Applications 37.4 (2010): 3024-3035.
[22] Wan, Jiafu, et al. (2016) "Mobile crowd sensing for traffic prediction in internet of vehicles." Sensors 16.1 (2016): 88.
[23] Permissions. Available online: https://www.google.com/permissions/products/ (accessed on 19 January 2019).
[24] Amin-Naseri, Mostafa, "Adopting and incorporating crowdsourced traffic data in advanced transportation management systems"(2018). Graduate Theses and Dissertations. 16541.
[25] Dixit V, Nair DJ, Chand S, LevinMW(2020) A simple crowdsourced delay-based traffic signal control. PLoS ONE 15(4): e0230598.
[26] He, Z., Guan, W., & Ma, S. (2013). A traffic-condition-based route guidance strategy for a single destination road network. Transportation Research Part C: Emerging Technologies, 32, 89-102.
[27] Bacon, J., Bejan, A., Beresford, A., Evans, D., Gibbens, R., & Moody, K. (2011). Using Real-Time Road Traffic Data to Evaluate Congestion. Dependable And Historic Computing, 93-117.
[28] Lu, N., Cheng, N., Zhang, N., Shen, X., & Mark, J. (2014). Connected Vehicles: Solutions and Challenges. IEEE Internet Of Things Journal, 1(4), 289-299.
[29] Karagiannis, G., Altintas, O., Ekici, E., Heijenk, G., Jarupan, B., Lin, K., & Weil, T. (2011). Vehicular Networking: A Survey and Tutorial on Requirements, Architectures, Challenges, Standards and Solutions. IEEE Communications Surveys & Tutorials, 13(4), 584-616.
[30] Jin, J., Ma, X., & Kosonen, I. (2017). An intelligent control system for traffic lights with simulation-based evaluation. Control Engineering Practice, 58, 24-33.
[31] Limited, IBI Consultanc. Best Practices for Traffic Signal Operations in India. New Delhi, India: Shakti Sustainable Energy Foundation, 2016. https://smartnet.niua.org/content/ffcc9b2e-c4c1-4fb3-9e67-93fe0a893e5e
[32] CoSiCoSt-WiTraC Compatible Adaptive Traffic Control System - C-DAC. (n.d.). Retrieved January 7, 2022, from https://www.cdac.in/index.aspx?id=pe_its_CoSiCoStBrochure
[33] New Delhi Traffic Report: Tomtom traffic index. report | TomTom Traffic Index. (n.d.). Retrieved January 7, 2022, from https://www.tomtom.com/en_gb/traffic-index/new-delhi-traffic/
[34] Nair DJ, Saxena N, GIlles F, Wijayaratna K, Dixit V. c: An Alternative to Conventional Speed Measurements [Internet]. Rochester, NY: Social Science Research Network; 2019 Jan. Report No.: ID 3325616.
[35] Nair DJ, Gilles F, Chand S, Saxena N, Dixit V (2019) Characterizing multicity urban traffic conditions using crowdsourced data. PLoS ONE 14 (3): e0212845.
[36] Hernández, J., Ossowski, S., & García-Serrano, A. (2002). Multiagent architectures for intelligent traffic management systems. Transportation Research Part C: Emerging Technologies, 10(5-6), 473-506.
[37] Logi, F., & Ritchie, S. (2002). A multi-agent architecture for cooperative inter-jurisdictional traffic congestion management. Transportation Research Part C: Emerging Technologies, 10(5-6), 507-527.
[38] Ding, J., Wang, C., Meng, F., & Wu, T. (2010). Real-time vehicle route guidance using vehicle-to-vehicle communication. IET Communications, 4(7), 870.
[39] Calabrese, F., Colonna, M., Lovisolo, P., Parata, D., & Ratti, C. (2011). Real-Time Urban Monitoring Using Cell Phones: A Case Study in Rome. IEEE Transactions On Intelligent Transportation Systems, 12(1), 141-151.
[40] He, Z., Cao, B., & Liu, Y. (2015). Accurate Real-Time Traffic Speed Estimation Using Infrastructure-Free Vehicular Networks. International Journal Of Distributed Sensor Networks, 2015, 1-19.
[41] Tostes, A. I. J., de LP Duarte-Figueiredo, F., Assunção, R., Salles, J., & Loureiro, A. A. (2013, August). From data to knowledge: city-wide traffic flows analysis and prediction using bing maps. In Proceedings of the 2nd ACM SIGKDD International Workshop on Urban Computing (p. 12). ACM.
[42] Juntunen, T., Kostakos, V., Perttunen, M., & Ferreira, D. (2012). Web tool for traffic engineers: direct manipulation and visualization of vehicular traffic using Google maps. MindTrek, 12, 209-210.
[43] Singh, B. S. R. B. J., & Xu, K. (2012). Real Time Prediction of Road Traffic Condition in London via Twitter and Related Sources by. Middlesex University.

[44] Kwak, D., Kim, D., Liu, R., Iftode, L., & Nath, B. (2014). Tweeting Traffic Image Reports on the Road. Proceedings Of The 6Th International Conference On Mobile Computing, Applications And Services.
[45] Mohan Rao, A., & Ramachandra Rao, K. (2012). MEASURING URBAN TRAFFIC CONGESTION – A REVIEW. International Journal For Traffic And Transport Engineering, 2(4), 286-305.
[46] Bing Maps Distance Matrix API. (n.d.). Retrieved November 11, 2018, from https://www.microsoft.com/en-us/maps/distance-matrix
[47] "Traffic Layer" | Google Maps Javascript API | Google Developers". Google Developers, 2017. Web. 27 Mar. 2017, https://developers.google.com/maps/documentation/javascript/examples/layer-traffic
[48] Xin, Wuping. A new architecture of adaptive traffic signal control in a data-rich environment. Diss. Polytechnic Institute of New York University, 2014.
[49] "Miller, Greg. "The Huge, Unseen Operation Behind The Accuracy Of Google Maps". WIRED, 2017. Web. 29 Mar. 2017, https://www.wired.com/2014/12/google-maps-ground-truth/
[50] Stenovec, Tim. "Google Has Gotten Incredibly Good At Predicting Traffic — Here's How". Business Insider, 2017. Web. 29 Mar. 2017, http://www.businessinsider.com/how-google-maps-knows-about-traffic-2015-11?IR=T
[51] Krichene, W., Drighès, B., & Bayen, A. M. (2015). Online learning of nash equilibria in congestion games. *SIAM Journal on Control and Optimization*, *53*(2), 1056-1081.
[52] Chiu, D., & Jain, R. (1989). Analysis of the increase and decrease algorithms for congestion avoidance in computer networks. Computer Networks And ISDN Systems, 17(1), 1-14.
[53] Zakariya, A. Y., & Rabia, S. I. (2016). Estimating the minimum delay optimal cycle length based on a time-dependent delay formula. Alexandria Engineering Journal, 55(3), 2509-2514.
[54] Koonce, P., Rodegerdts, L., Lee, K., Quayle, S., Beaird, S., Braud, C. & Urbanik, T. (2008). Traffic signal timing manual (No. FHWA-HOP-08-024).
[55] Bajpai, A., & Mathew, T. V. Development of an Interface between Signal Controller and Traffic Simulator, https://sumo.dlr.de/pdf/CTRG_Interface-SUMO.pdf
[56] Mishra, S., Bhattacharya, D., & Gupta, A. (2018). Congestion Adaptive Traffic Light Control and Notification Architecture Using Google Maps APIs. Data, 3(4), 67.
[57] TNM Staff Follow @thenewsminute. (2019, August 25). Hyderabad traffic dept to use Google Maps' real-time data to manage signals. Retrieved from https://www.thenewsminute.com/article/hyderabad-traffic-dept-use-google-maps-real-time-data-manage-signals-107797
[58] Mishra, S., Kushwaha, A., Aggrawal, D., & Gupta, A. (2019). Comparative emission study by real-time congestion monitoring for stable pollution policy on temporal and meso-spatial regions in Delhi. Journal of Cleaner Production, 224, 465-478.
[59] Deng, W., Xu, J., Gao, X. Z., & Zhao, H. (2020). An enhanced MSIQDE algorithm with novel multiple strategies for global optimization problems. IEEE Transactions on Systems, Man, and Cybernetics: Systems.
[60] Deng, W., Xu, J., Zhao, H., & Song, Y. (2020). A novel gate resource allocation method using improved PSO-based QEA. IEEE Transactions on Intelligent Transportation Systems.
[61] Rasheed, F., Yau, K. L. A., Noor, R. M., Wu, C., & Low, Y. C. (2020). Deep reinforcement learning for traffic signal control: A review. IEEE Access, 8, 208016-208044.
[62] Sakata, N., Fujimoto, K., & Maruta, I. (2021). On trajectory tracking control of simple port-Hamiltonian systems based on passivity based sliding mode control. IFAC-PapersOnLine, 54(19), 38-43.